# Review of Clustering-Based Recommender Systems

**Beregovskaya Irina[1] and Koroteev Mikhail[2,*]**

[1] Financial University under the government of Russian Federation; beregovskaya97@gmail.com
[2] Financial University under the government of Russian Federation; mvkoroteev@fa.ru
* Correspondence: mvkoroteev@fa.ru; Tel.: +7-999-849-12-36

**Abstract:** Recommender systems are one of the most applied methods in machine learning and find applications in many areas, ranging from economics to the Internet of things. This article provides a general overview of modern approaches to recommender system design using clustering as a preliminary step to improve overall performance. Using clustering can address several known issues in recommendation systems, including increasing the diversity, consistency, and reliability of recommendations; the data sparsity of user-preference matrices; and changes in user preferences over time.

This work will be useful for both beginners in the field of recommender systems and specialists in related fields that are interested in examining the applicability of recommender systems. This review is focused on the analysis of the scientific literature on the topics of recommender systems and clustering models that have appeared in recent years and contains a representative list of the literature for the further exploration of this topic. In the first part, a brief introduction to the so-called classic or traditional recommendation algorithms is given, along with an overview of the clustering problem.

In today's digital world, users suffer from the problem of information overload, and recommender systems are widely used as a decision support tool to solve this problem. Although recommender systems are a proven and affordable tool, the need to improve their recommendation ability and effectiveness is high. Among the various mechanisms available for generating similarity-based recommendations, collaborative filtering approaches are widely used. In addition to this approach, content-based filtering algorithms and hybrid filtering algorithms that combine the features of the first two varieties can be found. To improve the process of creating recommendations for various approaches, clustering methods are used with the aim of grouping users and increasing the accuracy of the recommendation system.

## 1. Introduction

Recommender systems have become quite common and are used in various fields [56–63]. With the development of Internet technologies, the flow of data from all areas leads to the problem of information overload. To solve this problem, many major websites and e-commerce sites use various convenient and effective recommendation systems to improve their quality of service and to attract and retain loyal users. For example, Amazon book recommendations, marketplace apps, YouTube videos, and Internet search results.

Tian et al. [1], for example, developed a personalized recommendation system for college libraries based on a hybrid recommendation algorithm, and discussed this topic in their article. The article raises the problem that, every year, the number of books in libraries increase, and users need to spend a great deal of time choosing the right book. At the same time, many books are not organized very effectively, which leads to unnecessary costs for libraries. These phenomena are caused by "information overload" and a library needs to rely on an information filtering mechanism to solve this problem. The information filtering mechanism is divided into two types: a search engine and a recommendation engine. The first mechanism uses a keyword to help users quickly find a suitable book and the second automatically recommends books to users. Personal recommendation systems seek to predict preferences based on interests, behavior, or other information from the user. Personalized recommendations can not only meet a user's needs, but can also help users to explore and discover new hobbies. The application of recommendation systems in university libraries solves the problem of book selection and increases the utilization of library resources.

There are three main categories of recommendation algorithms: collaborative filtering, content-based filtering, and hybrid recommendations.

1. Collaborative filtering is based on collecting and analyzing a large amount of information about the behavior, activities, and preferences of users and the predicting what a user likes based on the similarity of the user to other users.

2. The content-based filtering algorithm is based on a description of the element and a profile of a user's preferences. These algorithms try to recommend items that are similar to those that a user has liked in the past.

3. The hybrid recommendation algorithm combines collaborative filtering and content-based filtering. In some cases, hybrid approaches can be more effective.

*1.1. Collaborative Filtering Algorithm*

Collaborative filtering is the most widely used approach in terms of recommendations for providing services to users. The essence of this approach is to improve the ability of active users to find accurate and reliable neighbors. However, the collected data are extremely sparse in the custom item ranking matrix, and many of the existing similarity measurement methods used in collaborative filtering are not very efficient, which results in poor performance.

Collaborative filtering is a successful techniques in recommender systems, which recommends items to a user by analyzing the user's data; these data can be obtained by tracking browsing history, purchase records, rating records, etc.

Collaborative filtering (CF) does not use the content properties of items and can only search for similar users based on how users rated items. In a typical CF system, a user-item matrix is created in which a user's preference for an item is represented as a rating. CF estimates the similarity between a target user and other users, finds a neighborhood by selecting similar users, and then predicts the rating of each unrated item for the target user using the neighborhood ratings.

CF has the advantage that recommendations can only be made using ratings. This feature, however, also has some disadvantages: items that no one has rated cannot be recommended, and accurate recommendation results are difficult to obtain for users who have rated only a few items. In addition, a profile injection attack against CF (discussed in [21]) is another issue related to this feature. Attacking users or competing companies can insert fake user profiles into the user element matrix to influence predicted ratings, increasing the likelihood that their elements will be recommended or decreasing the likelihood that opponents' elements will be recommended.

CF algorithms are generally divided into memory-based and model-based collaborative filtering algorithms. In memory-based CFRSs, a custom member scoring matrix is built to generate appropriate recommendations, and the algorithm can also be further broken down into collaborative filtering based on users and members. The user-based CF algorithm computes the similarity between a target user and a neighboring user, and then the recommender system generates recommendations based on the interests of a highly-rated similar user. In CF, user-based recommendations are generated based on the assumption that a user with similar qualities to the target user in the present may have similar desires in the future. Likewise, the item-based collaborative filtering algorithm computes a similarity score between different items and provides recommendations to an active user. To make recommendations with CF based on items, item similarity is calculated with the assumption that items that are similar to previously consumed items may be purchased in the future. Model-based CF approaches are widely used to address data reduction and scalability issues through the use of a custom member rating database.

The user-based collaborative filtering (CF) algorithm is divided into three stages: creating a user model, finding the closest set of neighbors, and making recommendations.

In recommender systems, the user-member rating matrix (R) contains the ratings of m users for n items; U denotes a set of m users and I represents a set of n items. The rating data of the rating matrix are sparse, missing, or unknown rating data and are indicated by the symbol "?". $r_{ui}$

denotes the rating of user u for item i. Supposing that there are n users, $U = \{User_1, User_2, ... User_n\}$, and a set of item categories, $I = \{Item\_1, Item_2, ..., Item_m\}$, R is expressed by an N * M matrix where N is the number of users and M is the number of categories. The number of items selected by the user from category j is denoted as $R_{ij}$.

The user-based collaborative filtering recommendations system (UBCFRS) generates user-centric recommendations using an item-rating matrix, which is usually defined as usr × itm. In an item rating matrix, usr represents an active user with various items of interest, and itm denotes specific items in RS. When the target user wishes to receive an offer from the recommendation system, neighboring users with similar tastes to the target user are determined. Based on the assessment of the previous ratings of the items of neighboring users, an item that might be of interest to the target user is predicted. In other words, a product to be recommended to a customer is rated based on the preferences of neighboring users with similar qualities to the target user's.

The computational similarity method, which allows inferring the similarity between an active user and available users, plays an important role in the process of predicting the rating of a recommender system.

When ratings are explicitly presented, similarity can be easily determined using the Pearson's correlation coefficient (PCC) or Pearson's similarity metric (PSim), given that similar users tend to rate an item with similar rating points. Empirical analysis of different similarity measures relative to the CF recommender system shows that PSim performs better than other existing similarity measures when calculating relationships between users [50].

The ratings are predicted by an average approach using an aggregation function that calculates a kind of average of all neighboring users' ratings. Based on the calculated forecast, the set of elements with the highest rating is offered to the active user.

A user-based algorithm calculates the similarity between two users. Calculating the similarity between users is an important part of this approach. Similarity metrics used mostly include:

1. Cosine similarity: the cosine angle between the vectors is given by:

$$S_{u,v} = \frac{\sum_{i \in I_u \cap I_v} r_{u,i} * r_{v,i}}{\sqrt{\sum_{i \in I_u \cap I_v} r_{u,i}^2 \sum_{i \in I_u \cap I_v} r_{v,i}^2}} \quad (1)$$

2. Dot product: the cosine angle and magnitude of the vectors also matter.
3. Euclidian distance: the elementwise squared distance between two vectors.
4. Pearson similarity: is a coefficient given by:

$$r = \frac{\sum_{i=1}^{n}(x_i - \bar{x})(y_i - \bar{y})}{\sqrt{\sum_{i=1}^{n}(y_i - \bar{y})^2 \sum_{i=1}^{n}(y_i - \bar{y})^2}} \quad (2)$$

Following the nearest set of neighbors, $U_k$, a list of recommendations (B) is produced to make an offer to the target user $B_u$ = {item$_1$,item$_2$,...}.

Collaborative filtering often suffers from thinness problems. The user–item matrix can be very large and sparse, which complicates the performance of recommendations. The most active users will only read a small portion of the entire database. The sparseness of the matrix reaches 99.99% [1,12,33].

*1.2. Content-Based Filtering (CBF)*

Content-based recommendation systems [60,64,67] work in a different way. They assign a set of features (profile) to each user and each item. This profile is used to measure the similarity between users and items. These features usually come from a natural description of the object being recommended; for example, a movie profile typically contains information about its genre (action, comedy, etc.), cast, box office popularity, release date, etc. [65].

Thus, in order to build a CBF recommendation system we need to describe a set of features of an items. To directly compare user and item profiles, CBF heavily relies on similarity metrics—functions that compute how similar or different two feature vectors are.

CBF models do not compare users directly; they base their recommendations solely on the user's past behavior. They derive desired recommendations from the feature-based representation of the items in the database [66].

For the content-based recommender system algorithm, first, the features of items need to be defined. In a library system, for example, information about a book may include title, classification number, index number, author, publisher, price, keywords, title, and authors. These features are used in CBF models as item profile features.

A user's preference profile can be expressed as a set of n tuples W:

$$W_i = \{(w_1, v_1), (w_2, v_2), \ldots (w_n, v_n)\}, \tag{3}$$

where $w_i$ denotes the preference of user i and weights $v_n$ denote the importance of a feature to the user. Finally, various candidate items are compared with the user's previously read books, and the most appropriate books are recommended.

*1.3. Clustering Algorithms*

CF is a system that predicts what items should be recommended to target users based on ratings made by users who are similar to those target users. Therefore, we can expect an increase in forecasting accuracy due to the early grouping of similar users into the same cluster. If attacking user profiles are grouped into one cluster, predictions for other trusted users can be made without being affected by the attacks. On the other hand, if the profiles of the attacking users are similar to those of many trusted users, grouping users can increase the impact of the attacks. The main purpose of the clustering algorithm is to group similar users into one cluster. In clustering-based approaches, neighboring users from a cluster are selected for target users they approach.

Clustering is the task of grouping a set of objects so that objects in one cluster are more similar to each other than they are to those in other clusters [52]. Clustering is often used as an unsupervised machine-learning tool to find a hidden structure in large datasets. It is based on grouping items in a dataset into several groups, or clusters, such as items in the same group being, on average, more similar than they are to items in different groups. In clustering algorithms, each item in the whole dataset is considered as a point in n-dimensional space, where n is the number of features of the item.

One of the simplest and still most common clustering algorithms is k-means [53,54,55]. The idea of k-means is to define cluster centroids—a set point in n-dimensional space—so that each point is. The algorithm of the k-means method is described as follows:

Step 1. The set of k-means is determined as $m_1, \ldots, m_k$.

Step 2. Each observation is linked to a specific cluster, the average of which gives the least sum of squares within the cluster.

$$E = \sum_{i=1}^{k} \sum_{x \epsilon m_i} \|x \mu_i\|_2^2 \tag{4}$$

Step 3. The new mean of the centroids of the observations in the new clusters is calculated.

$$\mu_i = \frac{1}{|m_i|} \sum_{x \epsilon m_i} x \tag{5}$$

Step 4. The new centroids are compared with centroids calculated earlier; if there is a difference, go to Step 2, otherwise go to Step 5.

Step 5. Stop and display the result of the clusters.

In recommendation systems, similarity-based measures have traditionally been used to determine neighboring users for a target user. In real-time recommender systems, not all users can rate, are in interested in, or can familiarize themselves with all available items. When there is a relationship or interaction between a user and an item, the user–item rating matrix will be sparse. This critical issue affects the accuracy of rating predictions by the recommendation engine and is known as the sparsity problem. With the increasing need to solve the sparsity problem, but inability to do so, similarity-based models are inadequate for defining an effective list of similar users. In parallel, similarity measures are computationally complex, and using them as the data scale increases will lead to an exponential increase in complexity. To solve problems, such as similarity-based measures when selecting neighboring users, clustering techniques can be used to separate users into different clusters. Typically, clustering can be defined as the process of grouping or organizing users in a database into a cluster while maintaining a higher degree of similarity between them in that cluster. Hence, when a target user is found to be similar to a cluster of users, the user is then added to that cluster, and items of interest to the users of that particular cluster are recommended to the target user. Using clustering techniques in recommendation systems helps to identify groups of users with similar tastes, and this approach greatly improves performance by being immune to sparsity issues. Commonly used clustering techniques include fuzzy, self-organizing maps (SOM), and k-means clustering.

The combination of different clustering algorithms, or the same clustering algorithm with different settings, is known as a cluster ensemble (CE). Clustering ensembles can overcome the instability issues of autonomous clustering models.

## 2. Methods of Using Clustering to Improve the Quality of Recommendation Systems

### 2.1. Hybrid Filtering Algorithm for Recommendations

The authors of [1] considered hybrid recommender systems and highlighted their three main strategies. The first is to conduct separate collaborative and content-based filtering. The second is adding content-based filtering capabilities to collaborative filtering (or vice versa). The third is the combination of the previous two approaches into one model [2]. To reduce the sparseness in data, the authors applied k-means clustering before calculating the similarity.

In their experiments, the authors of [1] used a dataset from a university library. To combat the problem of sparsity, they replaced books with book categories, thus using a user–category matrix; and then carried out the clustering of users (k = 15). The sparsity of the matrix was calculated as the proportion of zero matrix elements among all matrix elements. The initial matrix sparseness was 99.99% and the authors managed to reduce it to 76.42%.

To compare the hybrid model of recommendations with conventional models, namely, CF and CBF (described above), the precision metric was chosen:

$$Precision = \frac{sum(R(u) \cap T(u))}{sum(R(u))}, \quad (6)$$

where R(u) denotes a recommendation sheet following a training set, and T(u) is a test case.

The authors of [1] ran a collaborative filtering algorithm, content-based filtering, and a hybrid algorithm for different sizes of training datasets. By increasing the size of the training sample, the accuracy metric gave a greater value for all algorithms, with the hybrid algorithm having much higher values than the rest.

The improved collaborative filtering algorithm solved the data sparseness problem by combining clustering algorithms. It can also effectively solve cold start problems when a new user or a new book, about which little is known, appears in the system.

The authors of [1] then used the Spark big data platform to improve the usability of the real-time model, thus creating a personalized recommendation system for university and college libraries. To some extent, this increases the efficiency of book recommendations and the number of books that users borrow, and reduces the wasteful use of university book resources.

### 2.2. Using Clustering to Increase Recommendation Diversity

The success of a recommendation algorithm is usually measured by its ability to accurately predict item ratings. There is no doubt that the accuracy of predictions is an important property of recommender algorithms. Much of the research on recommender systems has focused on improving accuracy (for example, see [69–73]); however, other factors play important roles in satisfying user needs. One such factor that has gained importance of late is the diversity of recommendation lists. For example, a system that offers movies to its users can be very accurate, that is, it can be very good at predicting user ratings by item; however, if a user's recommendation list consists of films of the same type (for example, only sci-fi films), it may not be very satisfying. A good system should also recommend a diverse set of films (films of different genres) to users.

There is, however, a trade-off between accuracy and diversity. That is, in most cases, diversity can only be increased at the expense of accuracy. Nevertheless, this decrease in accuracy may be preferable if user satisfaction increases. This is a well-known issue with recommendation systems [68].

It has been some time since recommender system researchers realized that predictive accuracy is not the only property that a successful recommender system must have. For example, McNee et al. (2006) argued that the assessment of recommender systems should go beyond the usual metrics of accuracy [4]. Herlocker et al. (2004) discussed novelty and insight as important parameters in evaluating recommender systems [5]. The concepts of novelty and insight are closely related to diversity, as increasing the diversity of the recommendation list increases the chances of recommending new and random items to the user.

Several strategies have been proposed to address the issue of diversity. In earlier studies, authors have proposed a greedy selection algorithm [6]. In this method, the items are first sorted according to their similarity to the target query, and then the algorithm begins to gradually build the search set (or recommendation list) so that both similarity and diversity are optimized. This is achieved in the following way: in the first iteration, the element that most closely resembles the target request is placed in the retrieval set, in the next iteration, the element that has the maximum combination of similarity with the target query and diversity concerning the retrieval that is already built is selected for the user's recommendation list. Iterations continue until the desired retrieval set size is reached. As noted in another article [7], this algorithm is highly inefficient, so those authors proposed a limited version of the greedy choice algorithm. In this version, the algorithm first selects the number (b) of elements closest to the target query, and then the greedy selection method is applied to that set of elements instead of the entire set of elements. As b approaches n (the number of elements), the complexity of this restricted version approaches the complexity of the greedy selection method. Zhang and Hurley [8,9] suggested another optimization-based approach based on the trade-off between similarity and diversity as a quadratic programming problem.

Aytekin and Karakaya [3] described a new method (called ClusDiv) that can be used to increase the diversity of lists of recommendations with a slight decrease in accuracy. The idea was to group items and build a list of recommendations by selecting items from different groups so that the diversity of recommendations is maximized without reducing the accuracy too much.

ClusDiv is applied after a prediction algorithm predicts unknown ratings of items offered to a user. Thus, real-world recommender systems can use ClusDiv without modifying existing prediction algorithms.

For a recommendation system to allow users to customize the diversity levels of their recommendation lists, the time complexity of the online recommendation algorithm must be very low. The time complexity of an algorithm is a measure of its computational efficiency relative to the growth of the dataset. More efficient algorithms may even take more time to process a certain amount of data, but the time needed increases when the amount of data increases. This makes them more efficient in the long run and far more scalable. ClusDiv has a very low time complexity, which makes it a highly scalable algorithm.

It also allows users to experiment with the recommendations provided by a system and to find a diverse set of items. ClusDiv includes a configurable parameter that allows users to customize the diversity level of their recommendation lists. They can adjust this setting independently of other users. Thus, it is up to users to decide how much they want to sacrifice accuracy in favor of diversity. It is still unclear how to implement this configuration in practice and the authors of [3] did not specify this.

No content information (such as genre or film director) about objects is required. Product rating information is enough to diversify recommendation lists.

To show the effectiveness of ClusDiv, the authors of [3] compared it only with the limited greedy method proposed by Smyth and McClave, since the other method proposed by Hurley and Zhang (2011) had similar levels of diversification efficiency and a slightly worse computational time complexity. As it turns out, ClusDiv was much faster than the restricted greedy method, while still providing a similar diversification efficiency.

Earlier studies have also used a cluster approach to better diversify featured products according to users' tastes. To do this, researchers grouped items in a user profile and recommended items that fit those individual clusters well rather than the entire user profile [10]. ClusDiv also organizes elements into groups; however, as described in detail below, it groups all the elements in a system, not just the elements in a user profile. That is, the goal of ClusDiv is not to recommend elements that suit users' tastes, but rather to recommend a diverse set of elements while maintaining the highest possible accuracy. This gives users the ability to encounter random items.

Ziegler et al. [11] defined a similarity metric based on classification taxonomies, according to which the similarity within a list was calculated. The authors of [3] proposed a heuristic algorithm for diversifying recommendation lists based on this similarity metric. As in other studies [68,69], the proposed method increased diversity but had some negative effects on accuracy. One of the important contributions of this work was to empirically show that overall user satisfaction increases with a variety of lists of recommendations. This result supports the claim that the accuracy of recommendation lists is not the only requirement for user satisfaction.

One possible metric for measuring the diversity of a user's recommendation list is calculated as the average difference of all pairs of items in the user's recommendation list. If I is the set of all elements, and U is the set of all users, then the diversity of the list of recommendations of a particular user, D (L (u)), can be defined as follows:

$$D(L(u)) = \frac{1}{N(N-1)} \sum_{i \epsilon R} \sum_{i \epsilon R, i \neq j} d(i,j), \qquad (7)$$

where $L(u) \in I$ is a list of user recommendations $u \in U$ and $N = |L(u)|$, and $d(i,j)$ is the dissimilarity of items $i, j \in I$, which is defined as one minus the similarity of items i and j.

Lists of recommendations with diversity values close to 1 will seem to be very diverse in appearance. In other words, the diversity value generated using Formula (5) will be dominated by the default values used for the missing estimates and will be misleading.

The authors chose to use z-scores for diversity values, which they called z-diversity, instead of using the absolute diversity values defined in (5). Formally, the z-diversity of the recommendation list is defined as:

$$ZD(L(u)) = \frac{D(L(u)) - D(I)}{SD(I)}, \qquad (8)$$

where I is the set of all elements in the dataset, and D(L(u)) and D(I) are the diversity of elements in L(u) and I, respectively. SD(I) is the standard deviation of the differences of all pairs in I.

Like many recommendation algorithms, ClusDiv has autonomous and online phases. In the autonomous phase, in addition to building a model, the authors constructed N (where N is the size of the list of recommendations, L(u)) clusters of elements, C = {$C_1$, $C_2$, ..., $C_n$}. Item clusters are built using the standard k-means clustering algorithm. Items are clustered based on their ratings, which are assigned by users. So the proposed approach is to cluster together items that were rated similarly by a large group of users. The premise is that items found in the same cluster are quite similar to all the users. Information about the contents of the items is not used. However, if information about the content of the items is available, and if the similarities between items based on this information can be determined, then those similarities can also be used when clustering items.

The ClusDiv algorithm is based on the construction of cluster weights (CW). CW is a matrix with a (u, i)th record, $CW_{ui}$, which contains the number of elements that cluster $C_i$ will add to the list of recommendations of user u. Users have their vector $CW_u$; thus, for example, if $CW_{ui} = 5$, then cluster $C_i$ will add five items to the list of recommendations of user u. It follows that the sum of the cluster weights for any user should be equal to N (the size of the list of recommendations).

After the authors generated the cluster weights of user u, they created a list of recommendations for u as follows: iterate over the items in the list of recommendations for u from top to bottom and move the item to the list of the first N for u if the weight of the cluster to which this element belongs to is greater than zero and subtract one from this cluster weight. Then, they continued to scan the list of recommendations in this way until all the cluster weights were equal to zero. When all the cluster weights were zero, the final list of recommendations was ready.

In the experiments, authors use three different recommender system algorithms: element-based, user-based collaborative filtering, and SVD (a variant of CF algorithm based on singular value decomposition of matrices, see [14]).

For all three datasets, ClusDiv's z-diversity and completeness performance were as good as the bounded greedy method (BG), which was designed primarily to optimize the diversity and completeness values in recommendation lists. The significant superiority of ClusDiv appeared when the issue of time complexity was considered. The authors also drew attention to the fact that the maximum level of diversity achieved by the BG method was higher than that of ClusDiv.

However, at high levels of diversity, the values of completeness were very low, which meant that the levels of diversity were useless in practice, as the recommendation lists would be very imprecise.

*2.3. CF with Clustering User Preferences*

The authors of [12] proposed a powerful new collaborative filtering algorithm based on clustering user preferences to reduce the impact of data sparsity. User groups were first introduced to differentiate between users with different preferences. Then, given the preferences of an active user, a set of nearest neighbors from the corresponding user group (or groups) was achieved. Additionally, a new similarity measurement method was proposed to calculate the similarity between users. Finally, experimental results on two sets of test data showed that the proposed algorithm was effective at improving the performance of recommender systems.

Developing recommendation technology can mainly be divided into two categories: a model-based approach and a memory-based approach [13]. The model-driven approach first builds a prediction model based on a custom member rating matrix and then predicts scores of the target members. Unlike the model-based approach, the memory-based approach first calculates the similarity between users/items, selects the top k similar users/items as active neighbors, and then generates predicted results. A memory-based approach can be divided into a user-based or element-based approach. In [12], the authors focused on improving the performance of custom recommender systems to reduce the impact of data sparsity.

Modifications and improvements to collaborative filtering are mainly found as two aspects: modification of the similarity measure and the choice of a user's neighbor when predicting a rating. Pearson's correlation coefficient (PCC) and cosine (COS) are often used as measures of similarity in recommender systems. Additionally, Jamali and Ester [15] proposed a modified PCC-based similarity measurement method using a sigmoid function (SPCC), which emphasizes the importance of common ranking elements. Intuitively, if users have more general rating elements, then they are more similar. According to the method of the cosine measure of similarity, the rating scale is not taken into account, and to solve the problem of shortage, an adjusted method of measuring cosine similarity (ACOS) was proposed [16].

In addition to the methods for measuring similarity suggested above, researchers also proposed many modified approaches for the selection of neighbors. For example, Kaleli [17] proposed an entropy-based optimization to generate a more qualified set of neighbors. It assigned a degree of uncertainty (DU) for each user and required neighbors with minimum differences in DU value and a maximum similarity value with the active user. Boumaza and Brun [18] introduced the concept of global neighbors, which are the neighbors of all active users. Kim and Yang [19] presented a threshold-based neighbor selection approach; in this approach, neighbors were determined in a certain range of choices based on the similarity of preferences. Anand and Bharadwaj [20] presented a recommendation framework combining both local and global similarities to address the problem of data sparsity, which allows to vary the importance given to global user similarity relative to local user similarity.

The authors of [13] presented an efficient collaborative filtering algorithm based on clustering user preferences that differ from those above. On the one hand, user groups are introduced to select more accurate and reliable neighbors for an active user. Users with different preferences have different rating habits. Thus, users can be combined into different user groups.

(1) An optimistic user group in which users prefer to rate high;
(2) A pessimistic group of users, in which users prefer to give low ratings;
(3) A neutral user group in which users tend to give reasonable ratings for products.

On the other hand, the authors noted that most of the previous similarity measurement methods were not suitable to account for user preference factors, and they proposed a new similarity measurement method for calculating the similarity between users in the clustering process. Moreover, extensive experiments showed that the algorithm proposed in [13] can significantly improve performance on sparse-rating data.

After calculating the similarity, k closest similar users are specified as the active user's neighbors, after which the prediction can be made for the target element. The recommended formula is defined as follows:

$$p_{ti} = \underline{r_t} + \frac{\sum_{u \epsilon U_{nei}} sim(t,u) * (r_{u,i} - \underline{r_u})}{\sum_{u \epsilon U_{nei}} |sim t, u|}, \quad (9)$$

where $p_{ti}$ denotes the forecast of active user t for target element i, $U_{nei}$ is the set of neighbors of active user t, $|U_{nei}|$ = k.

As discussed above, users can be divided into three different user groups. Suppose $C_o$, $C_p$, and $C_n$ represent the optimistic user group, the pessimistic user group, and the neutral user group, respectively. Meanwhile, $c_o$ is the clustering center $C_o$, $c_p$ is the clustering center $C_p$, and $c_n$ is the clustering center $C_n$.

In the process of clustering, the rating information of clustering centers has special characteristics; that is, $c_o$ prefers to give high marks, and the determination of user preferences depends on the similarity between the user and these clustering centers. Hence, an effective method of measuring similarity is useful for distributing the remaining users into different user groups. To emphasize the importance of user preference, the authors proposed a new similarity measurement method for calculating the similarity between users, as shown below:

$$sim(a,b)^{UPS} = exp\,exp\left(-\frac{\sum_{i \epsilon I_{ab}} |r_{a,i} - r_{b,i}|}{|I_{ab}|} * |\underline{r_a} - \underline{r_b}|\right) * \frac{|I_a| \cap |I_b|}{|I_a| \cup |I_b|} \quad (10)$$

Next, the authors developed an appropriate algorithm to make recommendations to an active user. They first calculated the similarity between users using the method they proposed, and the similarity matrix was denoted as $sim^{UPS}$. Then $c_o$, $c_p$, and $c_{n\,are}$ were defined as clustering centers with different preferences, respectively. Finally, users were categorized into different user groups based on their similarities. This generated various user groups, which were an optimistic user group $U_o$, a pessimistic user group $U_p$, and a neutral user group $U_n$. After completing the clustering process, the k nearest neighbors for the active user could be determined.

After obtaining a set of neighbors, $U_{nei}$, for active user t, one can predict the rating ($p_{ti}$) as follows:

$$p_{ti} = \underline{r_t} + \frac{\sum_{u \epsilon U_{nei}} sim^{UPS}(t,u) * (r_{u,i} - \underline{r_u})}{\sum_{u \epsilon U_{nei}} |sim^{UPS} t, u|} \quad (11)$$

To evaluate the performance of the algorithm proposed by the authors, time complexity analysis was required. The choice of clustering centers required additional time—O (m), where m denotes the number of users, and when the authors calculated the similarity between users of the proposed method, the computational complexity was O (m (m + 2)).

In [13], this algorithm was tested on two well-known datasets: MovieLens (ML, [74]) and HetRec2011–MovieLens (HRML, [75]).

To assess the performance of their proposed method, they used the mean square error (MAE) to measure the quality of the predictions, as well as the accuracy and completeness to measure the quality of a set of recommendations.

Over the course of experiments on the two data sets, it was revealed that, with an increase in the number of considered neighbors for an active user, the MAE indicator decreased.

When comparing the results of COS-CF and modified-COS-CF, the authors were convinced that the accuracy of COS-CF recommendations was lower than that of modified-COS-CF with an increase in the number of nearest neighbors. Likewise, modified-PCC-CF also clearly outperformed traditional PCC-based collaborative filtering (PCC-CF).

All the modified approaches had a higher recommendation accuracy than traditional algorithms.

This approach is based on the assumption that users have different rating habits. To distinguish between different typical users, the main work in this article is to develop a structure for distributing users into groups of users with different preferences. Hence, neighboring users of the active user can be found to have consistent preferences. Traditional methods of measuring Pearson's correlation coefficient and cosine similarity have drawbacks. In [13] a new similarity measurement method to look at user preferences from a local and global perspective, respectively, was proposed. In the course of experiments, the authors evaluated the effectiveness of their proposed algorithm for improving the quality and performance of recommendations, respectively, and experimental results for the two sets of control data demonstrated that the proposed algorithm performed better than some modern recommendation algorithms. In short, the proposed algorithm was effective at improving the performance of recommender systems.

*2.4. Using Clustering to Improve Recommendation Reliability*

Collaborative filtering is widely used by online vendors and review sites to recommend items based on the ratings of many users. However, this method has several problems, and one of them is the presence of attacks aimed at distorting the predicted ratings of specific elements. The authors of [21] proposed a collaborative filtering technique that reduces the impact of attacks while maintaining or improving prediction accuracy by repeatedly applying clustering to target data and predicting ratings for unrated items within each cluster. In addition to this, the usefulness of the method was investigated using a scoring method that measured the error between actual user ratings and predicted ratings. Additionally, attack resistance was investigated by comparing pre- and post-attack prediction errors.

Collaborative filtering (CF), the subject of this article, is one of the representative techniques used in recommender systems. CF predicts the ratings of unrated items by assessing the similarity between users and calculating a target item's rating prediction for a target user based on observed ratings from similar users.

However, CF has a vulnerability to profile injection attacks [22], which intend to distort the results of recommendations. In CF-based recommender systems, the quality of the recommendation can be influenced by the introduction of multiple user profiles for attacks, in which specific items are deliberately rated high or low. Eliminating this defect is important to improve the reliability of recommender systems.

Because CF searches for users that are similar to the target user and recommends items that those users prefer, it is expected that the prediction accuracy can be improved by pre-clustering similar users. However, you can increase the impact of attacks if the cluster sizes are too small. Thus, [21] proposed a forecasting method that performs clustering.

The prediction method first divides all users, including attackers, into multiple clusters, calculates the centroid of the users in each cluster as a representative cluster point, and then clusters again using the representative points to connect the split clusters. The similarity between users in the same cluster is then calculated, and item ratings are predicted using user similarity and ratings suggested by similar users within the cluster.

Clustering is used to separate users in a recommendation system into similar groups. Users are first divided into clusters using k-means clustering, and then clustering is performed again using the centroid of users belonging to each cluster. Each element $c_j$ in cluster $C = (c_1, \ldots, c_j, \ldots, c_n)$ can be calculated as follows:

$$c_j = \frac{\sum_{i=1}^{m} R_{ij}}{m}, \qquad (12)$$

where m is the number of users in the cluster, and $R_{ij}$ is the rating of the i-th user to the j-th element.

The number of clusters was set to 20–100 for the first clustering and 2 for the second clustering. It is necessary to provide a certain number of clusters for the first clustering and to ensure that the cluster size grows for the second clustering.

The goal of [21] is to reduce the impact of attacks while maintaining or improving the prediction accuracy. CF-based prediction is performed on items that are rated by users, and forecast accuracy is assessed by measuring the errors between the actual user-assigned ratings and the predicted CF-based ratings. In addition, after measuring the errors before and after attacks, the resistance to attacks is analyzed by calculating the difference between the errors, before and after the attacks, which is equal to the change in the predicted CF estimates before and after the attacks. MAE is used as a measure of measurement error.

There are several types of attacks against CF-based recommender systems. Although three types of attacks have been tested in experiments, due to limitations, only a discussion of an average attack [23] is given in this article. An average attack is carried out through attack user profiles, with ratings of randomly selected items around the average of each selected item and with ratings of targeted items within the highest or lowest rating.

According to the purpose of the attack, an attack aimed at increasing the popularity of a target is called a push attack, and an attack aimed at reducing the popularity of a target is called a nuclear attack. The authors of [21] focused on a push attack aimed at increasing the ranking of certain items. When performing a medium push attack by injecting attack user profiles into the source data, targets were randomly fetched and given the highest scores, and the other items, excluding the target items, were randomly selected to average the user ratings of the corresponding item. In the experiments, the number of attack users (attack size) and the number of randomly selected elements, except for the target elements (placeholder size), were changed to check the impact of attacks and the reliability of the CF recommendation in detail.

For the experiments, the authors of [21] used the well-known Movielens100K dataset [74]. In the experiments, CF-based prediction was performed for each method—no clustering, single clustering, and double clustering—and the errors between predicted ratings and actual user ratings were measured. The expected result was that prediction accuracy improves as the number of clusters increases.

Fake user profiles with a medium attack were then added and the error between the predicted ratings and the actual user ratings is measured. The trend was similar to the results before the attacks; that is, the error based on the one-shot clustering method was the smallest, followed by clustering twice, and then without clustering.

Focusing on the double clustering method, the average error difference was always less than at least one of the other two methods, and sometimes, it was the smallest among all three methods for some cases with certain attack sizes and non-target elements. This indicates that by specifying the appropriate number of clusters, the double clustering method can outperform the other two methods in terms of resistance to medium attacks.

Thus, [21] proposed a robust co-filtering method by running the clustering process and the rating forecasting process twice within clusters. Additionally, a method was proposed for assessing resilience by measuring errors between predicted ratings and actual ratings, before and after attacks, and calculating the difference between errors to investigate the impact of attacks. The experiments in [21] showed that a prediction method that performs clustering twice is effective in mitigating attacks.

*2.5. Using Clustering in Recommendation Systems to Reflect User Interest Change over Time*

CF algorithm's advantage is that it does not impose special requirements on the recommended types of resources and can work with unstructured complex objects [25]. However, with the ever-increasing number of users and resources of an e-commerce website, the traditional collaborative filtering recommendation algorithm is faced with problems of data sparseness, real-time change, extensibility, and so on. Therefore, it is difficult to ensure the required quality of a recommendation system.

To solve these problems, many scientists have carried out intensive research and have obtained some achievements. For example, based on the traditional method of measuring similarity [26], an improved method for calculating similarity has been proposed, which increases the recommended accuracy; the data sparseness problem was also effectively solved when matrix factorization methods, such as single value decomposition (SVD, [27]), non-negative matrix factorization (NMF, [28]), etc. They were applied in the joint filtering algorithm, and the real-time system was improved when clustering was introduced into the joint filtering algorithm. In the literature [29], the k-means method is used to cluster users and proposed projects, which reduces the cost of searching for the nearest neighbor.

The authors of [31] also presented additional attributes of projects proposed for users that have been assessed in the clustering process, in combination with user ratings and project attributes, user-clustering better reflects user interests and clustering results become more reliable. However, the algorithm does not take into account the situation where the interests of users can change over time, and the clustering of users cannot reflect the changing interests of users very well, and thus the problem of a new project (cold start) cannot be solved.

The authors of [31] used the temporal fade function to display user interests and change them multidimensionally, simultaneously with the introduction of the attributes of proposed projects, and presented an improved collaborative filtering algorithm based on user clustering.

Here, the joint filtering algorithm can be divided into three stages: data presentation, nearest neighbor search, and acquiring recommended results. The accuracy of the choice of the nearest neighbor, to a certain extent, determines the quality of the recommendation algorithm; that is, the method of measuring the similarity for the joint filtering algorithm is very important.

Currently, the similarity measurement method for the joint filtering algorithm is usually implemented in three ways [32]: vector cosine similarity, corrected cosine similarity, and Pearson correlation similarity.

The current user's rating for unrated projects can be predicted based on the current user's nearest neighbor rating information:

$$p_{ti} = \underline{r_t} + \frac{\sum_{u \in U_{nei}} sim(t,u) * (r_{u,i} - \underline{r_u})}{\sum_{u \in U_{nei}} |sim(t,u)|}, \quad (13)$$

Regarding the clustering of users, while there is no general clustering algorithm that can cluster different data, different applications have different clustering algorithms. The k-means clustering algorithm is simple and efficient, is suitable for large datasets, and can be very well implemented into the collaborative filtering algorithm; the authors of [31] chose it as the clustering algorithm in their work.

The steps to implement the authors' improved collaborative filtering algorithm based on clustering can be divided into two phases: choosing a recommended set of candidates for a project and an online Top-N recommendation.

Step 1. Selecting the recommended set of project candidates

Similar users are located in the same cluster by clustering users. The cluster in which the users are located is the set of candidates for the nearest neighbor search. The choice of the set of candidates recommended for the project should be based on the results of clustering the user and the project to make the recommended set of candidates for the project perfect and reliable; the following steps are needed:

- Supposing that the cluster in which user u is $C_u$, for $\forall u_i \in C_u$ its vector of interest $(Di)_{1r\,An}$ needs to be constructed.
- An improved method of calculating the degree of similarity.

$$sim(i \text{ is used}; j) = \frac{\sum_{a_x \in A}(R_{i,a_x} - \underline{R_i}) * (R_{j,a_x} - \underline{R_j})}{\sqrt{\sum_{a_x \in A}(R_{i,a_x} - \underline{R_i})^2 \sum_{a_x 0 A}(R_{j,a_x} - \underline{R_j})^2}}, \quad (14)$$

where A is a set of project attributes; $R_{i,a_x}, R_{j,a_x}$, respectively, represent the rating weights of user $u_i, u_j$ by project attribute $a_x$; to calculate the similarity for $\forall u_i \in C_u$, it is necessary to select users $K_u$ with the highest similarities as the nearest neighbors, and this is written as $C_{nei_u}$.

- It is necessary to take a rating set of projects $I_u$ from $R_{mn}$ according to $C_{uk}$ and u.
- $i_j \in I_u$, need to be used to find cluster $C_i j$, to which it belongs, for $\forall i_j \in C_{i_j}$ the vector attributes of the project need to be built.
- An improved method for calculating the degree of similarity needs to be used.

$$sim(i; j) = \frac{\sum_{a_x \in A}(A_{ii} A_{ji})}{\sqrt{\sum_{a_x \in A}(A_{ii})^2 \sum_{a_x \in A}(A_{ji})^2}}, \quad (15)$$

where A is a set of project attributes; $a_x$ is a project attribute; $A_{ii}, A_{ji}$, respectively, represent whether the project includes the $i_i, i_j$ attribute $a_x$ This is used to calculate the degree of similarity for $\forall i_j \in C_{i_j}$. It is necessary to select the projects $K_i$ with the greatest similarity, which will be the nearest neighbors and is written as $C_{nei_{i_j}}$.

- Calculate the union $C_{nei_i} = C_{nei_{i_1}} \cup C_{nei_{i_2}} \cup ... \cup C_{nei_{i_n}}$.
- It is necessary to delete projects in $C_{nei_i}$ that are rated by user u and compare the similarity; then, one must select projects $K_r$ with the highest degree of similarity, which will represent the project of user u recommended by the set of candidates and written as $W_u$.

Step 2. Online Top-N recommendation.

To obtain a recommended result for user u, we also need to predict the rating of projects in $W_u$ and obtain a Top-N recommendation. According to the ranking forecast for the recruitment of recommended candidate projects, the N highest-rated projects to be included in the recruitment are selected, thus completing the Top-N recommendation process.

Furthermore, the authors of the article in question used the MovieLens dataset in their experiments. The MAE (mean squared error) was used to assess the rating prediction errors; the authors used the recall rate and precision rate to assess the accuracy of the recommendation sheet.

After implementing the rating calculation procedure, the authors compared the MAE errors for different algorithms: the traditional joint filtering algorithm based on the user's similarity level using Pearson's correlation (P); a collaborative filtering algorithm based on combining user similarity calculation methods (Pearson with Salton) (PS); NMF algorithm; -c error MAE for the authors' proposed improved collaborative filtering algorithm (ICCFRA). The result showed that, compared to the P, PS, and NMF algorithms, ICCFRA sharply reduced the MAE, which significantly increased the quality of the rating forecast.

The accuracy of the ICCFRA algorithm, when generating recommendations, was the highest with recommendation lengths of 30, 40, and 50.

Thus, the execution time of the online algorithm was reduced by improving the real-time collaborative filtering algorithm. The experiment result for the MovieLens dataset shows that the algorithm significantly improved the MAE, as well as the recall rate and precision rate. In addition, the clustering-based collaborative filtering algorithm proposed in this article processes the original score matrix first using the time decreasing function, which solves the problem of the relevance of the original score.

*2.6. Using Clustering to Deal with Data Sparsity*

In practice, the effectiveness of CF models, as we have already seen, is limited by the sparseness of the rating matrix of historical users and the cold start of new users [34,35]. The sparseness of data indicates that historical users only rate a few items; for example, an audience, on average, and far fewer users leave comments (ratings) and view less than 2% of movies on a movie website. With an increase in historical data, the situation will be even more severe. The scarcity of rating data leads to a serious decrease in accuracy and causes the high computational cost of CF-based methods. A cold start means it is difficult to predict the preferences of new users who have no item records.

Researchers have proposed several CF best practices to overcome the above-mentioned limitations and improve the performance of the recommendation system. One class of a wide range of solutions is to take advantage of clustering or dimensionality reduction to eliminate the effect of historical sparseness in user ratings. Typical representatives of these methods are bicluster algorithms, singular value decomposition, the factorization of a non-negative matrix, etc. [36, 37, 38], and the key idea of these methods is to use local dense and low-dimensional modules of a rating matrix instead of the original sparse data in user ratings to assess the similarity between new users and historical users; they can then make recommendations using an improved similarity measure.

Another strategy for solving constraints in CF is to use some advanced similarity measures to improve the perception of sparse data and complex information. The traditional measures of similarity in CF, as we have seen in previous sections, are Pearson's correlation or cosine correlation.

The work of [33] presents a method of joint filtering based on biclustering and information entropy (CBE-CF) to overcome data sparseness and heterogeneity. Specifically, it takes advantage of biclustering to determine dense modules of a rating matrix and then measure the similarity between a new user and the dense modules based on a measure of information entropy. Finally, a linearly weighted combination of user-based CFs with an improved similarity measure and item-based CFs are used to fulfill the recommendation.

Although a user-based CF is widely used in various applications, the computational costs of measuring user similarity increase dramatically with an increase in the number of past users; consequently, the element-based CF is designed to adapt rapid response requirements to a large-scale product offering data to users. Unlike user-based CF, item-based CF first constructs a measure of item similarity, based on the common users, because the number of items is often much smaller than the number of users in most applications; this strategy can effectively reduce the computational cost of determining the k-nearest neighbors.

It is worth noting that, in practice, some users often share a common preference for certain group elements, the patterns of which can be well described by the consistency of local preferences among both the users (rows) and elements (columns) of a rating matrix, and are often used to address the sparseness of data in a recommendation system. The authors of [33] used biclustering techniques to identify combination patterns consisting of a local dense rating area for identified items with specific users. The general idea of biclustering is to iteratively aggregate the rows and columns of a rating matrix until convergence [39]. Specifically, for rating matrix R, X represents users (rows) and Y represents items (columns), and then I ∈ X and J ∈ Y indicate an indexed subset of users and items in the same cluster.

Information entropy, which is used in [33] to measure the similarity of a new user and dense modules, is a measure of the distribution of information of a random variable [33]; a high entropy means a tendency towards a uniform distribution, and conversely, a low entropy indicates a sharp distribution of the random variable.

Collaborative filtering performance can decrease as the number of items in the training dataset increases. In [33], the authors proposed a new collaborative filtering (CBE-CF) method for extracting local dense rating units to cope with data sparseness and the computational efficiency of traditional recommendation algorithms by introducing information entropy and biclustering in collaborative filtering. Experimental analysis shows the characteristics of the CBE-CF method proposed in [33] and the accuracy and computational costs are higher and lower than modern results on a set of reference data.

The CBE-CF recommender system method can be described in the following steps:

Step 1: Bicluster analysis is performed on the initial "user-element" rating matrix to determine its low-dimensional and dense local modules. Users in each specific cluster have identified item scoring templates, and each template points to a specific cluster.

Step 2: The informational entropy for each cluster obtained in Step 1 is calculated. In detail, the authors first count the number of elements assigned the same rating in a particular cluster and then estimate the probability that each rating for the identified cluster will be found. The entropy information for each cluster is then calculated, which can be used to measure local similarity between new users and clusters.

Step 3: Implementation of a user-based collaborative filtering algorithm. First, the authors sort, in ascending order, the differences in information entropy between all clusters $E_{p_{clu_i}}$ and the new user $E_{p_{new_i}}$ with the measure $E_{p_{diff_i}} = |E_{p_{new_i}} - E_{p_{clu_i}}|$ and then the first N clusters associated with the smallest differences are selected as the nearest neighborhoods for building the recommendation system. This strategy can effectively reduce the computational costs of assessing similarity because it simply focuses on a few predefined clusters instead of real-time similarities between a huge number of new user pairs and historical users. The authors assume that $I_{new}$ elements of the new user can be divided into $I_k$, k = 1, 2, ..., N, a set of elements associated with the first N neighboring cluster. Then the similarity between the new user is determined for $u_{new_k}$ and the cluster $c_{k\ is}$. Finally, the recommendation for a new user, $u_{new_k}$, can be implemented by taking the weighted average of N first nearest neighbors.

Step 4: Combinatorial collaborative filtration (CBE-CF). CF primarily takes advantage of local patterns of historical users and significantly reduces the computational costs for large-scale training data; however, this method does not take into account general patterns of historical data. Hence, the authors present combinatorial collaborative filtering, integrating the advantages of biclustering and information entropy CF and traditional element-based CF linearly; this model also maintains a low computational complexity:

$$\widehat{p_{u_{new}J}} = \lambda p_{u_{new}J} + (1-\lambda) p_{u_{new}J}^{item}. \quad (16)$$

In general, the proposed method is CBE-CF and takes O (mn) + O (k) time in the training phase, where m, n, k are the user number, element number, and cluster number, respectively.

The experiments also used 10-fold cross-validation to evaluate the performance of the new method and the other compared methods, so each of the two datasets are evenly divided into 10 datasets and, in turn, the contents of the nine datasets were selected as the training dataset and the remaining dataset acted as a test suite. State-of-the-art user-based CF, element-based CFs, were used to assess the advantages and disadvantages of the new CBE-CF method. In addition, the number of nearest neighbors was set to 50 for all CFs based on KNN.

The performance of the new CBE-CF method and the four other compared methods was evaluated, and the accuracy and computational costs are compared using the HML and NF datasets. The CBE-CF method was run based on optimal parameters. The new method had the highest forecast accuracy and relatively low computational costs compared to all four presented methods. In particular, the performance of the new method was better than that of the probabilistic model (probabilistic latent semantic analysis, PLSA) and the non-negative matrix factorization (NMF) model with a relatively low cost. The obvious observation is that the time to compute the user-based CF increased rapidly with the increase in training data, while the new CBE-CF method was not sensitive to the amount of training data.

To test the sparse data capability of the new CBE-CF method, the authors randomly split the NF dataset into 10 datasets of different scales, and then executed CF methods at these different scales. Interestingly, the new CBE-CF method provided improved accuracy using an extended training set, indicating that the new method could overcome the effects of the sparseness of the training data. However, in addition to the new CBE-CF, two other robust methods (NMF and PLSA) showed high computational costs for a large training set.

The feasibility of this strategy was validated on two sets of benchmarks using four comparison methods. Notably, deep learning-based CF methods also provided excellent predictive capabilities, although they suffered from high computational costs and large training sample sizes [41,42].

### 2.7. Using Clustering Ensemble to Improve Consistency of Recommendations

Although many traditional clustering mechanisms are used to group users in modern research, to generate optimal recommendations, it is still necessary to study the use of clustering methods based on biological factors. The work in [50] introduced a new clustering ensemble based on biological principles by combining swarm intelligence and fuzzy clustering models for collaborative user filtering. These approaches were evaluated on real, large-scale Yelp and TripAdvisor datasets to check the accuracy and consistency of the recommendations using standard rating metrics.

There are many clustering approaches available in user-based CFRS to provide user-friendly guidance, such as k-means, fuzzy C-means, and the SOM method. However, algorithms with biological factors are not widely used for clustering users. In [50], an attempt was made to use a biological-based intelligent clustering approach in custom collaborative filtering.

The nature-inspired approach works better than traditional models, and their metaheuristics are specifically designed to handle complex real-world applications. Traditional approaches have failed to solve optimization problems, while biological metaheuristic algorithms are known for providing efficiently optimized solutions. For several large-scale applications, biological metaheuristic methods have been recognized as the best solution and have proven to be effective. To solve real-time global optimization problems, the development of hybrid biological methods for solving complex problems is very important. Swarm intelligence provides promising results for optimization problems and analytical data models, inheriting the characteristics of biological systems. Due to their proven effectiveness, intelligent swarm models have been actively studied, and the resulting solutions have opened the way for innovative ideas.

New clustering models based on swarm intelligence have improved clustering results which have been achieved through greater adaptability. Various fields, such as pattern recognition, big data, and recommender systems, are adapting swarm intelligence-based clustering approaches to improve performance. In [50], a new smart swarm clustering ensemble model was developed for RS to address information overload.

The study in [50] presented stability as an additional metric for evaluating RS algorithms. A stability score is used to compute the consistency of the generated predictions of a target RS algorithm. The authors argued that similarity-based user clustering by leveraging swarm intelligence for the ensemble clustering method improves RS performance and yields better results at the expense of both accuracy and stability.

To overcome the limitations of conventional clustering algorithms, clustering models based on intelligent swarms, based on biological factors, have been introduced. Since swarm intelligence inherits biological traits and characteristics, it is useful for obtaining quality results for solving global optimization problems. For example, a hybrid clustering model optimizing a swarm of particles using C-means and k-means achieved improved clustering results compared to traditional models [51,52]. In this article, the authors present a hybrid clustering model through ensemble clustering using MWO and particle swarm optimization (PSO) with fuzzy models. The fuzzy clustering model computes the degree of membership in a cluster with other elements, while the hard clustering model maps each element to a specific cluster [50].

Metaheuristic optimization algorithms, such as GA (genetic algorithm), ACO (ant colony optimization), and PSO (particle swarm optimization), have solved many optimization problems [50]. PSO has become a generally accepted metaheuristic algorithm because of its simplicity and versatility, and it has been used as an important technique in various applications. In successful works, various clustering models with PSO have been proposed [50]. Many PSO-based hybrid clustering models have a proven clustering accuracy compared to traditional clustering approaches, such as k-means and fuzzy C-means. However, the PSO-based model requires the setting of parameters before being applied, and it is also relatively slower than the traditional clustering model, which is a noticeable disadvantage.

Several clustering models provide different results with the same dataset; as such, there is no universal clustering model for obtaining optimal solutions with different types of datasets. To solve the above problem, clustering ensemble (CE) is recognized as an effective approach [50]. The clustering ensemble combines different solutions of clustering algorithms, or combines the results of one clustering algorithm with different parameters to create a new and improved solution, which is usually defined as a consensus solution to a problem. A clustering ensemble can process distributed data and is capable of parallel processing. The main contribution of this article includes an overview of several clustering approaches for generating recommendations. A detailed description of existing clustering algorithms, such as k-means, C-means, PSO, and MWO, is presented to develop new user clustering algorithms. The authors also present a new CE method with swarm intelligence algorithms for clustering users to generate advanced recommendations.

In [50] a new recommendation system based on the biointensive cluster ensemble (BICE) was presented. The proposed BICE-based CFRS has three main segments: user clustering, prediction of user interests, and recommendation of generated travel suggestions.

The proposed BICE approach is designed to cluster users of a given dataset by using biological approaches and obtaining a final clustering result using a statistical ensemble model; the BICE-based CFRS then performs a neighborhood search of the active target user to include it in the appropriate cluster. Then, based on the current neighbors of the active target user in the cluster, ratings are estimated and a list of the first n recommendations is made, which is then presented to the user. The authors used two different approaches to predicting ratings: the average nearest neighbors approach and PSim.

The proposed CFRS setting is designed to generate BICE-based recommendations, and the same setting has been modified for other combinations of user-clustering-based recommendation approaches. Along with the BICE approach, the authors present three different combinations of hybrid user clustering approaches, HCE1, HCE2, and HCE3. The HCE1 approach is a combination of k-means, C-means, and K-PSO methods used to cluster users. The HCE2 approach corresponds to a combination of k-means, C-means, and FCM-PSO methods. The HCE3 approach is a hybrid combination of the k-means, C-means, and K-MWO methods.

Even though the BICE model proposed by the authors takes a little longer to generate recommendations, the resulting proposals turn out to be more accurate than using basic approaches.

The experimental results show that the proposed hybrid approaches are more efficient than existing stand-alone approaches. The proposed hybrid approaches perform well, both in terms of assessing accuracy and in terms of stability. The ensemble-clustering model of the BICE approach using K-PSO, FCM-PSO, and K-MWO generated effective user clusters [50].

## 3. Discussion

From this review, we can conclude that, in general, algorithms for recommender systems evolve and become more complicated, as in any field of machine learning. There is a trend for using hybrid approaches, assembling different models of the same type to improve performance and by combining models for different purposes in pipelines. As shown above, clustering can be quite effective as a preemptive stage before recommendation systems. However, the overall effectiveness depends now on both the recommender algorithm and the clustering model. This can lead to difficulties in creating, testing, and implementing these models in practice. As the authors of [1] wrote, "Hybrid approaches, making content-based and collaborative-based predictions separately and combining them could be more effective in books recommender systems. … Obviously, a hybrid algorithm based on the collaborative filtering algorithm and content-based algorithm improved the efficiency and quality of the recommendation algorithm. Meanwhile, it can also solve item cold start issues effectively." Furthermore, we see increasing usage of deep learning methods to derive inner latent representations of users and item profiles to deal with the massive degree of data sparseness. In addition, the impact of these methods is characterized as "dramatic performance improvements brought by deep learning" [41]. Similar to other fields of machine learning, the more data that are collected, the more complicated and deep models become regarding the use of these data.

Studies have shown that accuracy can no longer remain as the main efficiency metric of a recommender system. Users and businesses need to not only match their existing preferences, sealing them, and putting users into bubbles, but also to encourage exploration and diversity. This can be also related to the old "cold start" recommendation problem. When we focus on inherent product features, analyzing them more rigorously, we can achieve more desirable results that are not captured by simple accuracy measurements. As was shown using ClassDiv, enabling intellectual data preprocessing can help here significantly, "…it has been recognized that accurate prediction of rating values is not the only requirement for achieving user satisfaction. One other requirement, which has gained importance recently, is the diversity of recommendation lists. Being able to recommend a diverse set of items is important for user satisfaction since it gives the user a richer set of items to choose from and increases the chance of discovering new items."[3]. We acknowledge there is more to this problem than just clustering. We hope to see more elaborate research on data analysis for better recommendations soon, using, for example, new emerging text understanding tools based on deep language models, such as BERT or GPT.

Another issue is the different rating habits of users. There is always inconsistency in user preferences, and effective recommender systems need to consider these issues. Using preemptive clustering to distinguish different groups of users may be promising, as shown in the studies above: "Our approach is based on an assumption that users have different rating habits. For distinguishing different typical users, the primary work in this paper is to design a framework to assign users into user groups with different preferences. Therefore, the neighbor users of the active user can be found with consistent preference. … To solve this problem, we proposed a new similarity measure method to consider user preference from the local and global perspectives respectively. In addition, an example was illustrated in our paper, which has proved that the proposed similarity measure method is more effective and suitable for calculating the similarity between users." [12].

As was shown in [21], clustering also can improve recommendation robustness by eliminating the possibility to perform specific attacks on the recommender system by constructing an artificial user profile to manipulate the output of the algorithm. "However, there are several problems with this method, and one of them is the existence of attacks that intend to distort the predicted ratings of specific items." [21]. This is very relevant to the current trend in machine learning for the exploration of fairness, robustness, and reliability of black-box machine learning methods used for decision-making support [76]. As intelligent systems gain popularity in every aspect of economic and social life, even more attention will and should be devoted to investigating different ways to ensure their abilities to withstand intentional attacks and inherent biases in training datasets.

Another interesting issue with enterprise recommendation systems is how to take into account constant changes in user preferences and behavior. As was noted in [24], "The traditional collaborative filtering recommendation algorithm based on user rating is very sparse, without because the user changes over time, not a good predictor of user interest, and the nearest neighbor query range is not conducive to a real-time recommendation, for the project problem is not a good solution.". Traditional recommender systems, both CF and CBF-based, simply do not have any concept of time within them. This may be an issue if users are present in commercial systems long enough to manifest significant changes in behavior. We suppose that this can have a major effect on a timescale of several years on average, though significant changes can appear very quickly in the very beginning of a user's experience within a certain system due to forming new consumption habits [77]. Thus, these improvements and new results and methods can be useful, not only to those who build long-lasting online services, but potentially everyone who uses recommendation systems to capture dynamic user interests.